\documentclass[english,preprint]{article}
\usepackage[utf8]{inputenc}
\usepackage[margin=1in,footskip=0.25in]{geometry}
\usepackage{authblk}
\usepackage{comment}
\usepackage{hyperref}
\usepackage{amsmath}
\usepackage{amssymb}
\usepackage{graphicx}
\usepackage{wrapfig}
\usepackage{multicol}
\usepackage{multirow}
\usepackage[super]{nth}
\usepackage[backend=biber, uniquename=init, giveninits=true, style=apa]{biblatex}
\usepackage{listings}
\usepackage{subfig}
\usepackage{xcolor}
\usepackage[bottom]{footmisc}

\DeclareNameAlias{sortname}{family-given}

\definecolor{pyblue}{RGB}{31, 119, 180}
\definecolor{pyorange}{RGB}{255, 127, 14}
\definecolor{pygreen}{RGB}{44, 160, 44}
\definecolor{pyred}{RGB}{214, 39, 40}
\definecolor{pypurple}{RGB}{148, 103, 189}
\definecolor{pybrown}{RGB}{140, 86, 75}
\definecolor{pypink}{RGB}{227, 119, 194}

\title{The Potential of Ridesharing Adoption and its Effects on CO2 Emissions and Customer Experience}

\author{Maximilian Kaufmann}
\author{Jan Nagler}

\affil{Deep Dynamics Group, Frankfurt School of Finance \& Management, Frankfurt, Germany\\
Centre for Human and Machine Intelligence, Frankfurt School of Finance \& Management, Frankfurt, Germany}

\begin{document}
\maketitle
\begin{abstract}
Taxi services are an integral part of urban transport and are a major contributor to air pollution and traffic congestion, which adversely affect human life and health. Sharing taxi rides is one way to 
reduce the unfavorable effects of cab services on cities. However, this comes at the expense of passenger discomfort, 
quantified 
in terms of longer travel times. Taxi ridesharing is a sophisticated mode of urban transport that combines individual trip requests with similar spatiotemporal characteristics into a shared ride. 
We propose a one-to-one sharing strategy 
that pairs trips with similar starting and ending points. 
We examine the method using an open dataset with trip information on over 165 million taxi rides. 
We show that the cumulative journey time can be reduced by 48 percent while maintaining a relatively low level of passenger inconvenience, with a total average delay compared to an individual mobility case of 6 minutes and 42 seconds. This advantage is accompanied by decreases in emissions of 20.129 tons on an ordinary day and a potential fare reduction of 49 percent, which could point to a widespread passenger acceptance of shared taxi services. 
Overall, a matching rate of 13 percent is reached while  
a 27 percent matching rate is attained for 
high-demand areas. Compared to many-to-many sharing dynamic routing methodologies, 
our scheme 
is easier to implement and operate, making fewer assumptions about data availability and customer acceptance.    
\end{abstract}

\section{Introduction}\label{Introduction}
One of the biggest problems facing cities worldwide is vehicular traffic congestion and the resulting air pollution. It has a high price in terms of money and life \parencite{santi2014quantifying}.\par

The widespread use of private cars to meet daily mobility needs is one of the most significant issues with managing the planet's finite resources. In addition to being utterly inefficient, it involves massive unneeded material transport, excessive energy waste, urban traffic jams, air pollution, and the exploitation of human resources for driving cars rather than engaging in paid employment or leisure activities \parencite{herminghaus2019mean}.\par

Meanwhile, the World Health Organization estimates that outdoor air pollution is one of the significant risks for non-communicable diseases, which is primarily brought on by vehicular traffic and is responsible for over three million deaths per year worldwide. 
About ninety percent of people living in urban areas are exposed to levels of fine particulate matter that exceed WHO Air Quality Guidelines. As New York City constitutes the basis of the cab trips referred to in this research, we need to mention that air pollution has significantly decreased in many high-income countries, including the US, because of efforts to reduce smog-forming emissions as well as particulate matter \parencite{WHStatistics2016}.\par

Another major problem that almost every populous city must contend with is congested roads, from which, in addition to higher pollution levels, economic losses result \parencite{sweet2011does}. According to a study from the Partnership of New York City, the annual cost of traffic congestion in the New York metropolitan area for 2018 alone is estimated to be 
20 Billion USD \parencite{ConPricing2018}. To curb this issue to some extent, The City Council of New York City is considering introducing a tolling program that would price every trip into southern Manhattan at up to 
23 USD  \parencite{ManhattanNYTToll}.\par

However, with the recent advancements in communication technologies, new business models and approaches could be installed that have the potential to address the problems of road congestion and air pollution in urban areas. For instance, using real-time data creates new opportunities to exploit spare capacity and enables unprecedented monitoring of the urban mobility infrastructure. These developments open up the potential to create new, more innovative transportation systems based on the sharing of cars and effectively provide services that could replace old public transportation infrastructure with on-demand taxi sharing. The advantage of the widespread use of smartphones and their ability to run real-time applications is another driver for implementing ridesharing services \parencite{shaheen2019shared}. UberPool or Lyft Shared Rides provides
examples for 
successfully 
implementing ridesharing services, which may 
raise hopes for a more efficient future.\par

On the other hand, ridesharing can also come with disadvantages. The overall service and waiting times may increase, and passengers may be concerned about the reliability of the shared ride. Moreover, privacy and security issues may be related to traveling with strangers in a shared vehicle. Lastly, the potential positive environmental effects must be treated with caution, as the reduced cost of taxi rides could also lead to increased demand or other rebound effects (\cite{lopez2014smarter}, \cite{storch2021incentive}).\par

Here, we analyze 
the potential of taxi ridesharing in New York City 
using an approach that could be installed in 
real-life cases. 
We also address
the effects of such an approach on the environment and ride quality.
\par

The structure of our paper unfolds
as follows.
Chapter \ref{Data Profiling} outlines the data structure and substance. 
The central part is presented in Chapter \ref{Analysis of the Ridesharing Potential}, where the one-to-one ridesharing approach is portrayed, and the empirical findings are described. 
Chapter \ref{Discussion} discusses the main findings and how a ridesharing provider may benefit.

\subsubsection*{Ridesharing}
When using ridesharing, two or more trip requests with similar origins and destinations are combined to transport multiple passengers simultaneously in one 
vehicle (\parencite{alonso2017demand}). By combining different trips into one shared ride, ridesharing increases the average number of people per vehicle, decreases the number of vehicles needed to meet the exact demand, and reduces traffic and other adverse effects of urban mobility on the environment \parencite{merlin2019transportation}.\par

Taxi ridesharing, also known as shared taxis or collective taxis, is an innovative mode of public transport with flexible scheduling and routing that matches, in real-time, at least two different trip requests with similar spatiotemporal characteristics to a shared 
taxi \parencite{barann2017open}.\par

Earlier studies on what we today call ridesharing have focused on small-scale ride-pooling approaches, which are not based on analyzing taxi trip records but rather on surveys. One example is an analysis of \parencite{caulfield2009estimating} on one-day data of commuting trips reported as part of a census survey made in Dublin. Caulfield found that 4 \% of respondents carpooled to work. They estimated that this ridesharing saved 12,674 tons of CO2 emissions per year.

\subsubsection*{Dynamic Ridesharing}
The developments in information and communication technologies have not just benefitted businesses that could implement a better ridesharing approach using real-time data. It also allowed the analysis of large-scale data using a dynamic approach. Dynamic ridesharing allows ridesharing to occur at short notice and between strangers who do not know each other's itineraries. The greater flexibility of dynamic ridesharing provides additional opportunities to maximize the benefits of sharing and improve the system's 
efficiency \parencite{lokhandwala2018dynamic}.\par

In a dynamic ridesharing system, matching the appropriate drivers to form the shared ride is critical.
Therefore, many researchers are focused on developing algorithms for perfect ride-matching. These calculations get more complex the more dimensions someone considers \parencite{lokhandwala2018dynamic}. For this reason, one can see that the recent works \parencite{santi2014quantifying} differ from the more recent ones \parencite{alonso2017demand} regarding the number of ride requests pooled into one vehicle.\par

Santi and coworkers \parencite{santi2014quantifying} introduced the concept of shareability networks and proposed a mathematical model to quantify the benefits of ridesharing. This work is seen as a pioneer in dynamic ridesharing as its results and methodology refer to numerous newer research pieces (\cite{alonso2017demand}, \cite{lokhandwala2018dynamic}, \cite{barann2017open}). \parencite{santi2014quantifying} analyzed taxi trip data in New York City and concluded that ridesharing could reduce cumulative trip length by more than 40 \%. This model restricted the sharing to a maximum of two ride requests per driver, ignoring the potential benefits of a more flexible system. Moreover, it assumed that all riders' tolerance level for journey delays was the same for drivers and passengers.

\subsubsection*{Ridesharing with Autonomous Vehicles}
With the rapid development of autonomous vehicles, the interest in the potential of ridesharing using such automobiles has received significant 
attention (\cite{alonso2017demand},\cite{fagnant2018dynamic}). Unlike taxis, whose drivers need to change shifts and take breaks, autonomous vehicles can be available around the clock. Moreover, there is no need to consider the taxi drivers' preferences and waiting times. Therefore, shared autonomous taxis can offer additional benefits compared to traditional ridesharing. The high-capacity ridesharing model of autonomous vehicles
by \parencite{alonso2017demand} can satisfy up to 99 \% of requests, with an average total delay of 2.5 minutes, by reducing the active operating taxis by 25 \%.

\subsubsection*{
Main Idea of Static Ridesharing}
Although the implementation of
a flexible and dynamic method has the potential to optimally address the problem of sharing taxi rides, these systems reveal inconveniences in real life scenarios
because customers might be reluctant to accept picking up and dropping off multiple passengers, almost all of them strangers, during a shared
ride \parencite{barann2017open}.\par

As mentioned, the many-to-many approach relies on complex matching algorithms, such as optimal routing, combining, and rerouting. Therefore, it could impose operational challenges for taxi operators \parencite{barann2017open}.\par

An excellent example of a recent study focusing on ridesharing using a non-dynamic approach was created by \parencite{barann2017open}. As an example of large-class
spatial sharing problems, this work 
proposes a framework that enables the analysis of fundamental trade-offs between the advantages and drawbacks of taxi ridesharing systems at the city level. Dividing taxi trips into clusters based on their pickup and drop-off destinations and grouping after a spatiotemporal constraint of matched trips, the research indicated that 48\% of taxi trips could be shared. The underlying assumption made implies 
that every passenger 
can walk to an assigned pickup destination if the distance is at most 500 meters. The drop-off of all passengers would happen in the same manner, and the targeted location must be reached on foot too. \parencite{barann2017open} argues that this model makes taxi ridesharing easier to implement. It reduces customers' perceived inconvenience by having all passengers meet at the pickup destination and letting them decide whether to share the ride with the other party \parencite{barann2017open}.\par

For the reasons named above, the benefits of a static approach may outweigh those of a dynamic approach.
Especially implementing a static model that considers a multi-ridesharing scenario 
may be easy 
to realize for a shared service provider. 
Therefore,
we employ the ridesharing potential using a one-to-one model, 
in contrast to \parencite{barann2017open}, by proposing 
a system that does not cluster the demands but uses a grid system of hexagons to define the pickup area. 
Moreover, our 
approach 
assumes smaller walking distances at the pickup locations and no walking distance at the drop-off location, assuming that a city walking difference of a maximum of 500 meters is unrealistic and drop-offs should be made at every target destination. Overall, this should limit the total delays. A time penalty for every drop-off is installed to account for a reasonable drop-off delay. 

\subsubsection*{Economic Incentives and Psychological Barriers}
Despite the numerous benefits of ridesharing, some challenges need to be addressed in the analysis to get a realistic understanding of the whole. One of the main challenges is understanding the factors that drive individuals to adopt ridesharing and how to effectively encourage more people to use shared rides instead of traditional modes of individual mobility. The current understanding of these conditions needs to be improved, making it challenging to implement effective strategies for promoting the widespread adoption of ridesharing \parencite{storch2021incentive}.\par

Recent research \parencite{storch2021incentive} 
suggests
that people considering ridesharing services weigh four primary incentives when deciding whether to request a single or shared ride: financial discounts, anticipated detours, the unknown length of the journey, and the inconvenience of riding in a car with strangers.\par

\parencite{storch2021incentive} predicts a sharp transition to high-sharing adoption for any given user preferences, implying that even a moderate increase in financial incentives or a minor improvement in service quality may disproportionately increase ridesharing adoption of user groups currently in the low-sharing regime under a variety of conditions.\par

In 
our analysis, these significant incentives are well considered. 
In particular, we study the 
relative potential of the fare reduction in a shared scenario 
and compare 
airport fares with total fares showed the possible financial incentives based on the pickup and drop-off locations.\par

To determine the resulting delays of this approach, the three primary variables assessing the trip delay are the size of the pickup and drop-off hexagon, the time interval or departure interval, and the drop-off rank. For every drop-off, a ridesharing participant can expect a delay of 20 seconds.\par

The inconvenience of riding with strangers can also be contained using this methodology. Every trip passenger, like in public transportation, can first assess with whom they would potentially share the trip and decide whether they want to do it. Compared to a dynamic approach, every ridesharing participant in the vehicle is not forced to share with whoever and whenever they wish \parencite{barann2017open}.\par

Finally, deciding on the proper drop-off resolution size, only journeys that concluded at a limited distance range were matched. 
As a result, 
we only consider 
trips that did not make lengthy detours to other places, and due to the small size of the drop-off destination, there was no need to add a distance constraint to reach comprehensive results.

\section{Data Profiling}\label{Data Profiling}
The dataset 
is accessible through NYC OpenData and contains 
pickup and drop-off geo-locations in the form of longitude and latitude and the respective trip dates and times, accurate to the second. Other variables include the passenger count, payment type, fare amount, extra costs, tax, airport fee, tolls, tips, and total amount in US-Dollars \parencite{taxitripdata2014}. The baseline of all initially considered variables is listed in Table \ref{tab:variables}.

\subsection{Data Availability}
In order to assess the ridesharing potential using the one-to-one approach, a publicly available dataset on yellow taxi trips from 2014, published by the NYC Taxi and Limousine Commission (TLC), is evaluated. The TLC provides data from 2009 to 2021, including several hundred million journeys conducted by yellow taxis, green taxis, and ride-hailing services. In recent years, the data supplier TLC has discontinued entering the precise geo-coordinates of pickup and drop-off points. Consequently, instead of working with a more recent dataset, the research must rely on outdated data from 2014, 2015,
or 2016 \parencite{taxitripdata2014}.\par

The dataset of 2014 consists 
of more than 165 million trips and a file size of 
25 gigabytes \parencite{taxitripdata2014}. The dataset of 2014 was preferred over the others because ride-hailing services like Uber and Lyft had yet to be extensively adopted to influence taxi ride demand at that time. Therefore, the 2014 trip data is more typical of the city's total demand than the 2015 and 2016 data \parencite{tlcfactbook2014}.\par

\begin{table}
\begin{center}
\begin{tabular}{ll}
\hline
\multicolumn{1}{|l|}{\textbf{Variable}} & \multicolumn{1}{l|}{\textbf{Description}} \\ \hline
Pickup Location                         & Longitude and Latitude                    \\
Pickup Datetime                         & Accurate to the Second                    \\
Drop-off Location                       & Longitude and Latitude                    \\
Drop-off Datetime                       & Accurate to the Second                    \\
Trip Distance in KM                     & Trip Distance in Miles * 1.609            \\
Passenger Count                         & Passengers excl. Taxi driver              \\
Fare Amount in USD                      & Total Fare Amount - Tip Amount            \\
Avg. Speed in Km/h                      & (Trip Distance / Trip Duration) * 3600    \\
Trip Duration in sec.                   & Drop-off Datetime - Pickup Datetime       \\
Great Circle Distance                   & Pickup vs. Drop-off Coordinates           \\ \hline
\end{tabular}
\caption{Variables}
\label{tab:variables}
\end{center}
\end{table}

\subsection{Data Cleaning}
\subsubsection{Geo Locations}
\begin{figure}
    \centering
    \includegraphics[width = 0.9\textwidth]{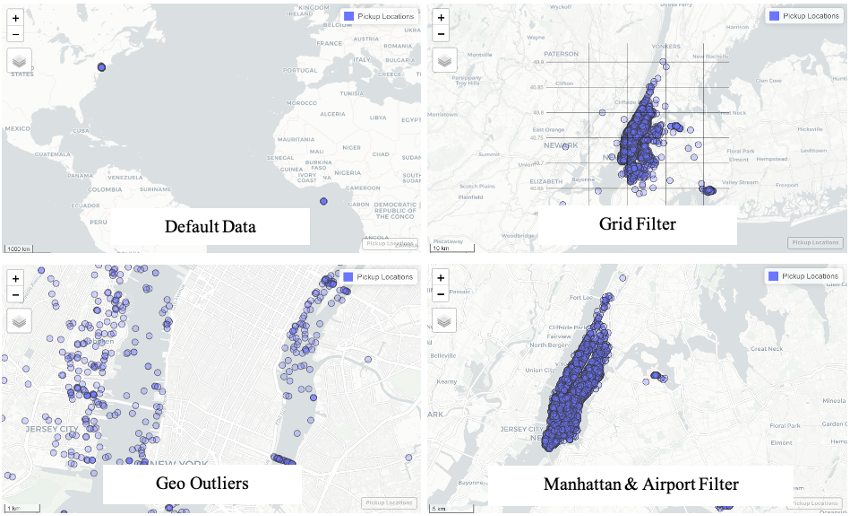}
    \caption{Geometric Locations and Outlier Detection}
    \label{fig:geo_locations}
\end{figure}

Severe outliers in the pickup and drop-off geo-locations were filtered first using the grid filter (see Figure \ref{fig:geo_locations}). Two percent of observations were omitted in this step. \par
As a next step, the data must be cleaned from outliers close to New York City but located in impossible locations that could not be addressed by the grid filter (see Figure \ref{fig:geo_locations}). A reason for these distortions could be uncontrolled biases resulting from urban canyons that may have slightly distorted GPS locations during data collection \parencite{santi2014quantifying}. Using a publicly available library called nycgeo, it is possible to filter for geographical and administrative borders in New York City \parencite{nycgeopackage}. However, the New York City filter resulted in a shallow loss of observations (see Figure 1). \par
The operational area of the yellow taxis is in Manhattan. Taxis outside of Manhattan but in New York City are called green Taxis and are not considered in this research. La Guardia and JFK airports are the only two dense pickup or drop-off destinations outside Manhattan operated with yellow taxis \parencite{tlcfactbook2014}. 
Visual inspection 
shows that the density of pickups and drop-offs outside Manhattan, apart from the airports, is decreasing dramatically. The last step separates Manhattan, La Guardia, and JFK Airport journeys from the rest. This step leads to a total data loss of 15 percent. 

\subsubsection{Trip Distance}
The great circle distance, which is the direct distance taking into account the spherical shape of the Earth, was used to compare all travel distances. The calculated distance was sometimes less than the actual distance traveled, 
which is geometrically impossible \parencite{donovan2017empirically}. Those observations were therefore excluded from the dataset. Additional cleaning procedures were carried out to eliminate irrational entries. \par
Trips less than 500 meters or more than 60 kilometers were eliminated too. Manual investigation revealed that those trips frequently had inflated times or distances. Additionally, all journeys with similar origins and destinations were abandoned by excluding all trip records with great circle distances 
below 100 meters \parencite{santi2014quantifying}. Trips with extraordinary
lengths, for instance, those that span less than a minute or longer than two hours, and trips with speeds below 5 km/h and above 88 km/h, which resembles the maximum New York State speed limitation of 55 mph were eliminated too (\cite{barann2017open},\cite{speedlimits}).\par

\subsubsection{Passenger Count}
Regarding the TLC \parencite{freqaskedquestionsnyc}, the maximum number of passengers allowed in a yellow cab taxi is four or five, excluding the driver, depending on whether the trip is operated with a four or five-person taxicab. However, additional passengers below the age of seven are allowed when held on an adult's lap. Hence, all trips with a passenger count above six will be excluded \parencite{costello2018big}.

\subsubsection{Trip Fares}
The research article \parencite{zhang2012smarter} is based on the 2009 yellow taxi trip data, showing that the distribution of fares closely resembles the distribution of trip length. This demonstrates that the primary determinant of the fare amount is trip distance. Nevertheless, \parencite{zhang2012smarter} also suggests separating pickups and drop-offs outside and inside the downtown area and considering rush hours that will affect the trip duration and, therefore, the fare amount.\par
\begin{figure}
    \centering
    \includegraphics[width = 0.9\textwidth]{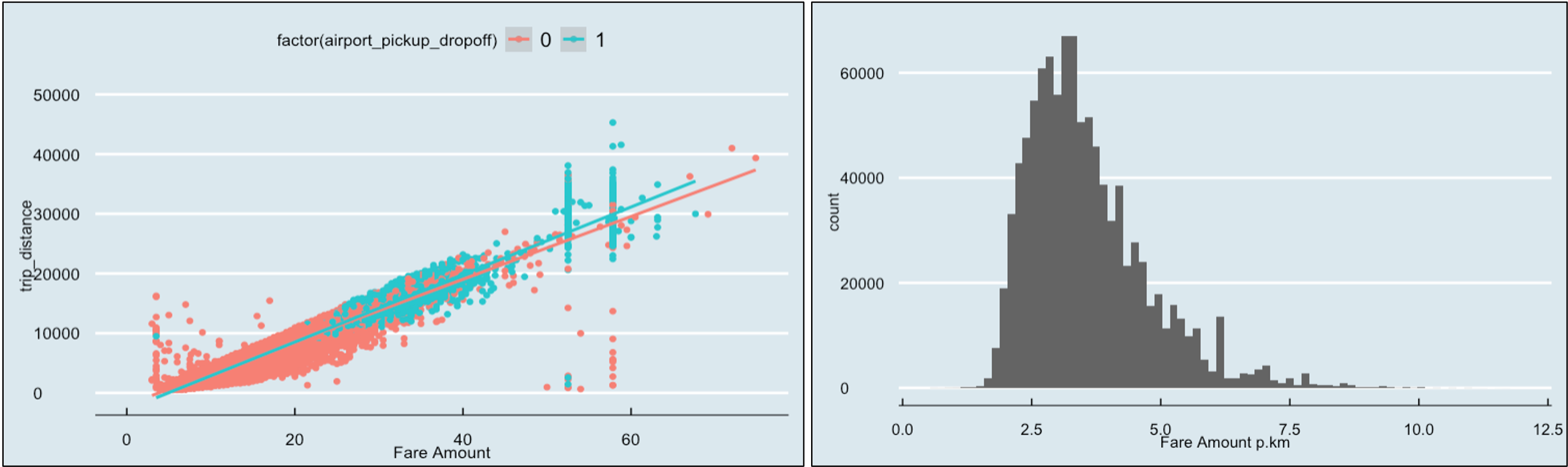}
    \caption{Trip Fare vs. Trip Distance}
    \label{fig:trip_fare_distance}
\end{figure}

When analyzing Figure \ref{fig:trip_fare_distance}, we 
assess that airport pickups and drop-offs usually come with the highest trip fare. The reason is that the two airports, JFK and La Guardia, are the only two locations outside the Manhattan borough. We can also assess a clear correlation between the fare amount and the trip distance. The histogram allows a visual assessment of outliers. It is based on the total ride fare per kilometer. All values below \$ 1.5 and above \$ 8.5 per kilometer were excluded from future analysis. 

\subsection{Data Exploration}
\begin{figure}
    \centering
    \includegraphics[width = 0.9\textwidth]{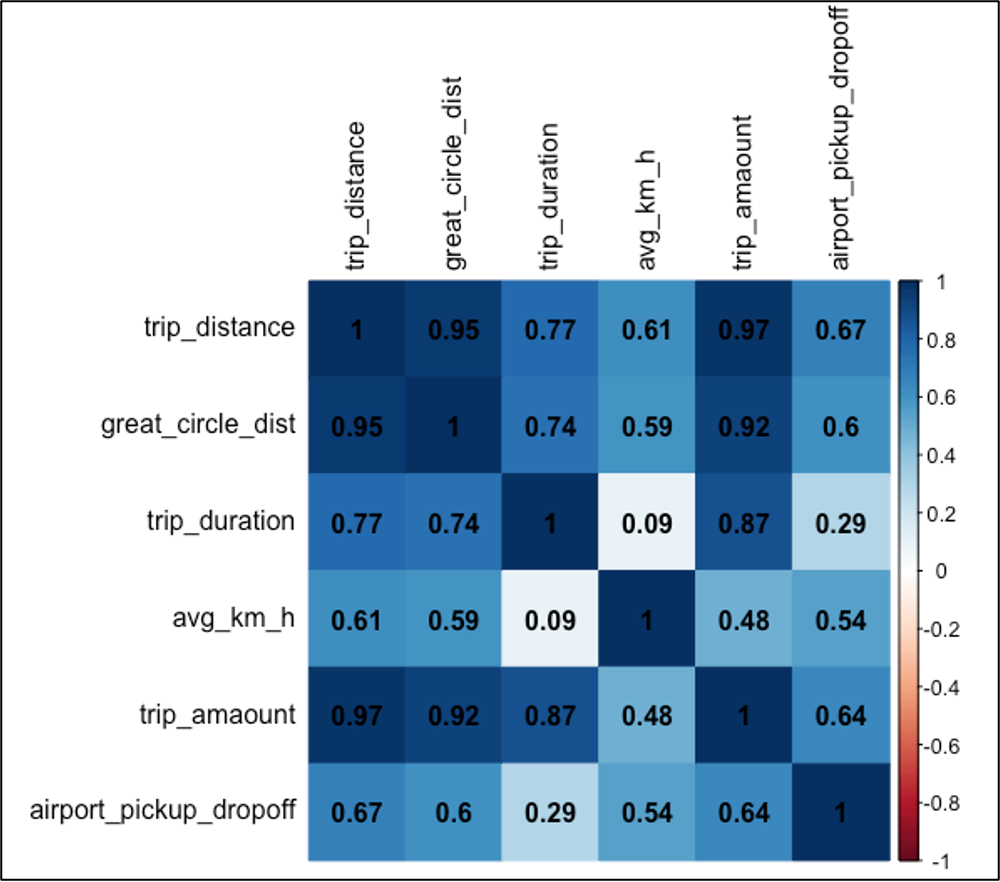}
    \caption{Correlation Matrix}
    \label{fig:correlation_matrix}
\end{figure}

Figure \ref{fig:correlation_matrix} confirms the naive expectation 
that 
fare  amount, 
trip duration 
and trip distance are highly correlated. 
Yet, 
the trip fare exhibits a higher correlation with 
the distance than the duration. 
This also aligns with an analysis by Zhang and coworkers 
\parencite{zhang2012smarter}. 
We further assume 
that the trip fare is even better approximated using the great circle distance instead of the trip duration. However, we must consider statistical measures to claim this statement with confidence. 
We 
also 
notice the positive correlation for 
the trips to and from the airport 
regarding 
trip length 
and 
travel costs.\par
\begin{figure}
    \centering
    \includegraphics[width = 0.9\textwidth]{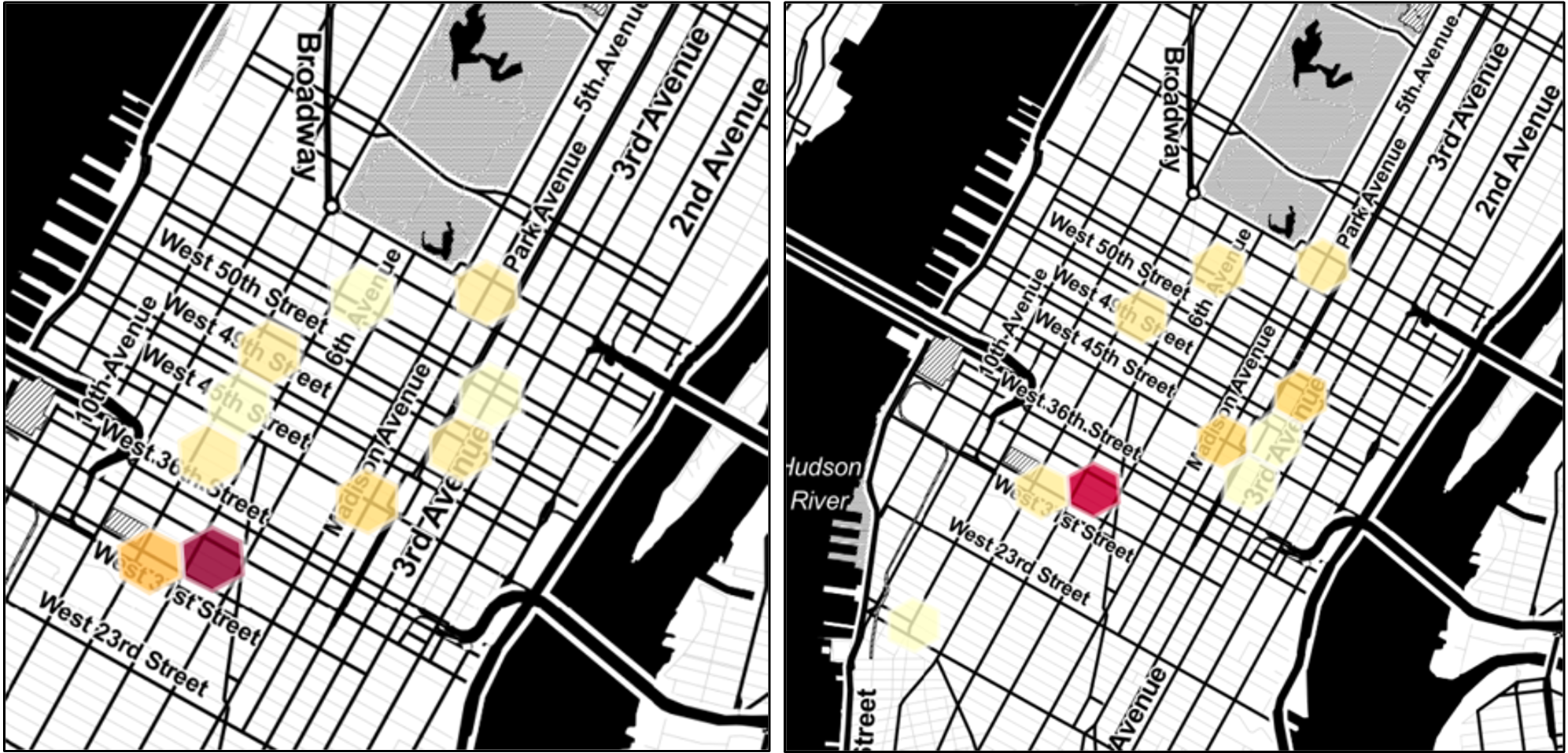}
    \caption{Busiest Pickup (left) and Drop-off (right) Locations}
    \label{fig:pickup_demands}
\end{figure}

The highest demand for taxi rides, represented by the number of pickups, is visualized in Figure \ref{fig:pickup_demands}. The darker a hexagon is colored, the more pickups or drop-offs there have been in the area. There is a precise centering of areas with the highest demand, as the top ten areas are all located in southern Manhattan. The two locations with the highest taxi demand are both at Madison Square Garden, with 340 thousand registered pickups in March. That area accounts for 2.75 \% of all recorded taxi trips of that month. The visualized ten hexagonal areas comprise 1.25 million pickups and represent 10 \% of all registered pickups in Manhattan. The most popular exit points (see Figure \ref{fig:pickup_demands} - right) display a similar picture. 330 thousand drop-offs are recorded at Madison Square Garden, and the top 10 drop-off destinations sum up to 1.2 million taxi trips representing 9.6 \% of all rides. Considering these figures raises hopes that a static approach can yield good results.\par
\begin{figure}
    \centering
    \includegraphics[width = 0.9\textwidth]{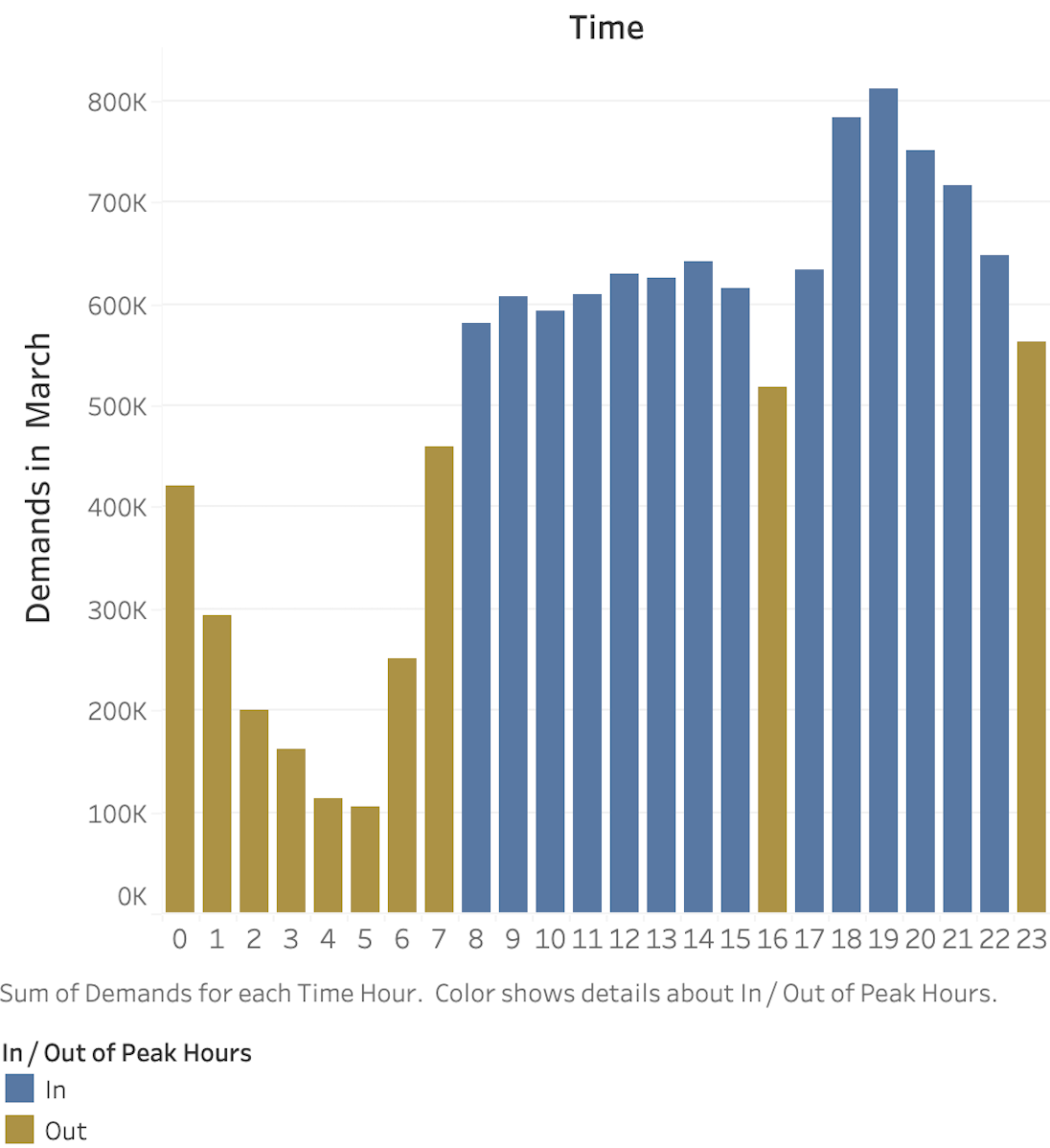}
    \caption{Peak Hours}
    \label{fig:peak_hours}
\end{figure}

In order to assess hours with the highest travel demands, the month of March is visualized in groups (see Figure \ref{fig:peak_hours}). Colored blue is the peak demand period; The morning-noon peak, from 08:00 to 15:00, and the evening peak, from 17:00 to 22:00, show the most rides per hour. This is important as we can later assess how the ridesharing potential reacts if solely implemented within these peak demand periods.\par

\subsubsection{Unused Capacity}
\begin{figure}
    \centering
    \includegraphics[width = 0.9\textwidth]{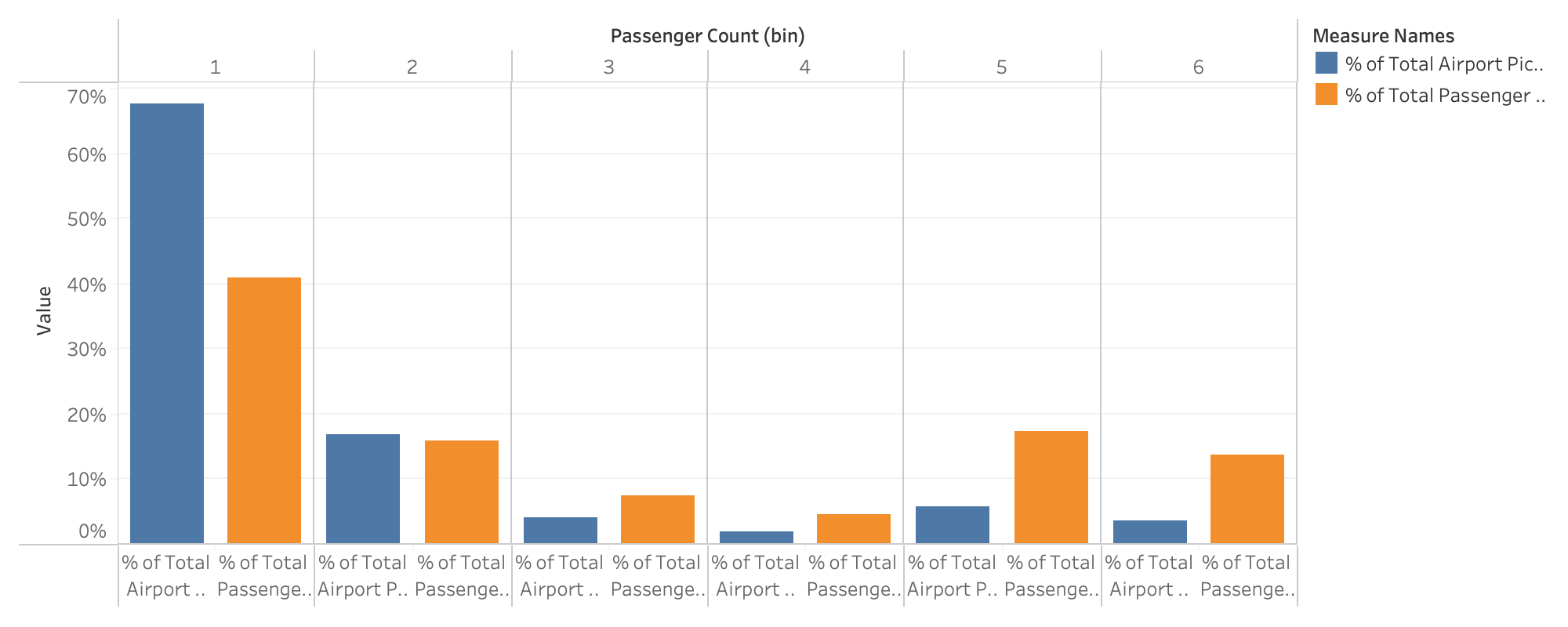}
    \caption{Occupancy - Total Trips vs. Airport Trips}
    \label{fig:unused_capacity}
\end{figure}

Inspection of 
Figure \ref{fig:unused_capacity} suggests that 
despite the highest taxi trip fares due to the longest average taxi journeys to and from the two airports (see Figure \ref{fig:trip_fare_distance}), the vehicle occupancy rates are low compared to the average occupancy in New York's Taxis. The only rides outside Manhattan we consider in this analysis due to their high request rate are JFK and LaGuardia airports. Only about 30\% of taxi trips to and from the airports have a load factor of more than one person, compared to the average total load factor of 65\%. Since the financial incentive should be highest for trips to and from the airports due to the higher travel costs, we will also examine this in the following.\par

\section{
Ridesharing Potential}
\label{Analysis of the Ridesharing Potential}
Here,  
using a simple static approach, we analyze the ridesharing potential for
densely populated cities. 
Our approach is based on
clustering pickup and drop-off points 
\parencite{barann2017open}, while also 
adopting the methodology of \parencite{liu2022exploring}
where 
the urban area is divided into hexagonal partitions.

With our analysis, 
we can evaluate the ridesharing potential without restricting ourselves in terms of car capacity \parencite{barann2017open}, whereas in a dynamic routing system proposed by (\cite{alonso2017demand},\cite{santi2014quantifying}), with an increasing number of potential rides that could be shared, the routing system gets mere complex as an increasing amount of nodes and therefore routes need to be taken into account.\par
The quality of service parameter is chosen as an absolute time rather than a relative increase of the travel time because it is consistent with similar realizations in the literature and is driven by the fact that absolute delay information is likely more helpful than the percent assessment of travel time increase for potential customers of a shared taxi service \parencite{santi2014quantifying}.

\subsection{
Model}\label{Mehodology 3.1}
\begin{figure}
    \centering
    \includegraphics[width = 0.28\textwidth, angle =-90]{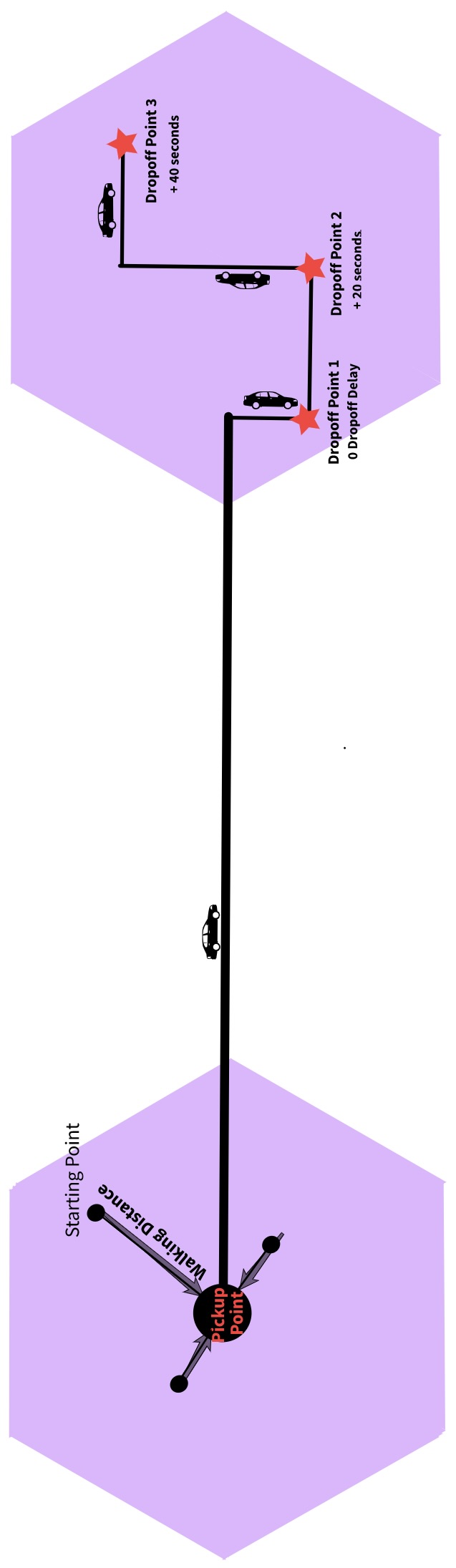}
    \caption{
    Methodology}
    \label{model_art}
\end{figure}

As a starting point, we divide the urban space into hexagonal partitions as L = \{l1, l2, li, ln\}. The duration of a day is discretized into time intervals as T = \{t1, t2, ti, tn\} 
which represent the time window at the pickup destination.\par
In order to create these hexagons latitude and longitude of pickup and drop-off points are converted to an H3 index, which is a string representation of the hexagon that the point falls in (\cite{h3source},\cite{h3library}). The data is then grouped by the pickup H3 indices, drop-off H3 indices, and pickup time within a given time window. 

Whenever the number of trips within a group exceeds 1, there is a demand for ridesharing. We assume all trips that fall into the defined group can be shared using one vehicle. As the number of ride requests and the accompanying passengers sometimes exceed the capacity of yellow taxis, we assume that at the determined pickup points, there is always the option to switch to a larger vehicle, such as a van.\par

Total carpooling potential can be determined by calculating the number of trips that start and end in the same H3 cell in a ridesharing scenario versus the number of trips in a non-ridesharing scenario. As a result, the ridesharing potential could be described as a ratio between the total trips in a ridesharing scenario and whole taxi trips.\par
However, similar to more intricate many-to-many Taxi Ride Sharing systems, this approach is constrained by several restrictions and other assumptions listed below. 

\subsubsection{Vehicle Constraints and Assumptions}
A regular yellow taxicab has a passenger limitation of only five persons \parencite{freqaskedquestionsnyc}. Most ridesharing situations can be fulfilled with this type of vehicle \parencite{barann2017open}. \par
Here, we allow for 
multi-ridesharing 
because this enables 
us to examine the full ridesharing potential. 
After all, a limitation can drastically reduce this potential \parencite{santi2014quantifying}. \par
In a real-life situation, multi-ridesharing that exceeds the 5-passenger limit can be performed using vans, minibusses, or large buses. Allowing multi-ridesharing, on the other hand, has a negative influence on waiting times \parencite{barann2017open}. However, passengers' inconveniences could be compensated with increased savings due to the greater total distance saved.

\subsubsection{Spatial Constraints and Assumptions}
The \textbf{starting point} of every participant in a shared ride scenario is the original pickup location of the ride request.\par
We assume that the new \textbf{pickup location} is always the center of a hexagon by neglecting that the exact point could be in an inaccessible location, for instance, in the middle of the street or a park. \par
The \textbf{pickup distance} is the great circle distance between the starting point and the pickup location. We assume every participant will walk this distance.\par
The \textbf{drop-off location} will be the exact location as in a non-ridesharing scenario.\par
The \textbf{total trip distance} in a shared scenario is the maximum trip distance of all ride requests or individual mobility trips that are pooled into a shared ride. This statement agrees with the assumption that the drop-offs do not result in any detours that could lengthen or shorten the trip. \par
\begin{figure}
    \centering
    \includegraphics[width = 0.9\textwidth]{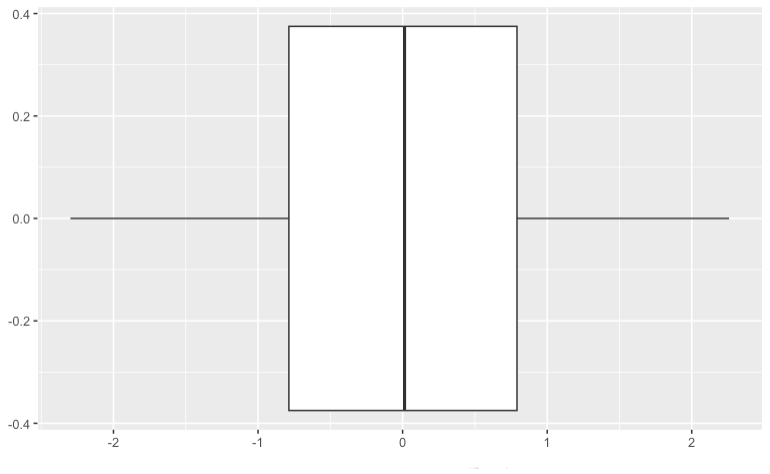}
    \caption{Great Circle Distance}
    \label{fig:great_circle}
\end{figure}

The \textbf{individual trip distance} will stay the same as in the original individual mobility case, as all participants are getting dropped off at their desired destination. However, as we relocate the pickup location to the hexagon center in a ridesharing scenario, the 
individual trip riding distance will change, depending on the starting point. If the starting point is closer to the drop-off location than the pickup point (hexagon center), the individual trip distance will increase and vice versa. By analyzing the great circle distance between all starting points and drop-offs in an individual mobility scenario and comparing it with the great circle distances in a shared scenario, we can observe that the difference in trip distance averages zero (see Figure \ref{fig:great_circle}). Based on this observation,
we assume that the individual trip distance for every ridesharing participant stays 
approximately 
the same as in an individual mobility scenario.

\subsubsection{Temporal Constraints and Assumptions}
The \textbf{starting time} of every participant in a shared ride scenario is the original pickup time of the ride request.\par
The \textbf{arrival time} is the walking duration to the previously defined pickup location added to the pickup starting time.\par
The \textbf{walking duration} is based on a study \parencite{amanda2006new} determining walking speed and attributes of pedestrians on lower Manhattan's sidewalks was observed. The median speed of all observed pedestrians was 4.67 kilometers per hour, which is slower than researchers found in previous studies (\cite{fruin1992designing},\cite{knoblauch1996field}) but could be explained by the fact that the majority of the observations took place in the middle of the day \parencite{amanda2006new}. 
Given that 
roughly 50\% of individuals in this study walked slower than the median speed,
some participants were likely not walking to a designated location, such as a taxi pickup place, and were not slowed down by distractions. Therefore, we use the 4.67 km/h to determine the walking duration. As a foundation, we use the pickup distance to approximate the walking distance and calculate the walking speed using the 4.67 km/h.\par
The \textbf{time window} or interval is the time difference between the trip announcement and departure time. Within this time window, we assume a vehicle with unlimited capacity is waiting at the previously defined pickup location. It applies that the number of daily intervals is 1440 minutes divided by the interval.\par
The \textbf{departure time} is constant and dependent on the size of the time window, but for every sharing scenario, the departure time must be greater or the same as the arrival time.\par
The \textbf{drop-off rank} is installed to approximate the individual delay for dropping off passengers as we assume the taxi drops everyone off at an individual location. This rank is based on the individual trip length in a shared taxi. The passenger with the shortest trip length is ranked first, and the passenger with the most extended trip length is ranked as the number of pooled taxi trips in the vehicle. By ranking every trip request in a shared scenario, we can penalize those with longer trip distances, as they are getting dropped off based on individual rank.\par
To specify the \textbf{drop-off delay}, we assume that every drop-off delays the total trip duration by 20 seconds. This delay is added to the remaining passengers in the vehicle and depends on the drop-off rank.\par
The \textbf{individual trip duration} will stay the same as in the original individual mobility case for the reasons described in the individual trip distance section.
\begin{itemize}
  \item \textit{Drop-off delay = 20 sec. * rank – 20sec}
  \item \textit{Total delay = walking duration + departure time – arrival time + drop-off delay}
  \item \textit{Drop-off time = Departure time + individual trip duration + drop-off delay}
\end{itemize}
Using the \textbf{great circle distance} between the starting point and drop-off location, we approximated the walking duration of the whole trip. Knowing the total trip walking duration,
we 
exclude every individual mobility scenario that reaches the drop-off point earlier when walking compared to participating in ridesharing.\par
Another constraint that must be installed is a fixed \textbf{drop-off delay} for every passenger in case the number of ride requests in one shared vehicle is too large. Too large means that the accumulated drop-off delay of all passengers would exceed the delay they would suffer if they get dropped off all at once and walk to the drop-off location.

\subsection{What-If Analysis and Robustness}\label{What-IF Analysis 3.2}
Previous studies show a positive correlation between ridesharing opportunities and both variables; time interval and cluster size \parencite{barann2017open}.\par
In general, a what-if analysis allows for the data-intensive modeling of complicated systems and analyses the results under alternative hypotheses \parencite{golfarelli2006designing}.
In short, the analysis helps to assess 
the robustness of parameter choices.

The analysis should help us to
study how resolution size and time interval will influence ridesharing opportunities. The following what-if study evaluates the ridesharing potential with changing parameters resolution and interval to quantify their impact on the outcomes for the proposed one-to-one ridesharing model (see Table \ref{tab:resolution_size_delay}). 
The analysis considered taxi trips between the 19th and 26th of March.\par
We consider time intervals from 1 to 25 and a pickup resolution of either 9 or 10. To add some flexibility while minimizing inconvenience for passengers, 
we
set the drop-off resolution to size 9, as the drop-offs are driven in the car. Therefore, a resolution of 10 would likely be too small and drastically influence the sharing potential. Whereas considering a resolution size of 8 is irrational as the side length of that hexagon is 530 meters and too extensive considering the presented drop-off methodology. 

\subsubsection{Hexagon Resolution}
\begin{figure}
    \centering
    \includegraphics[width = 0.9\textwidth]{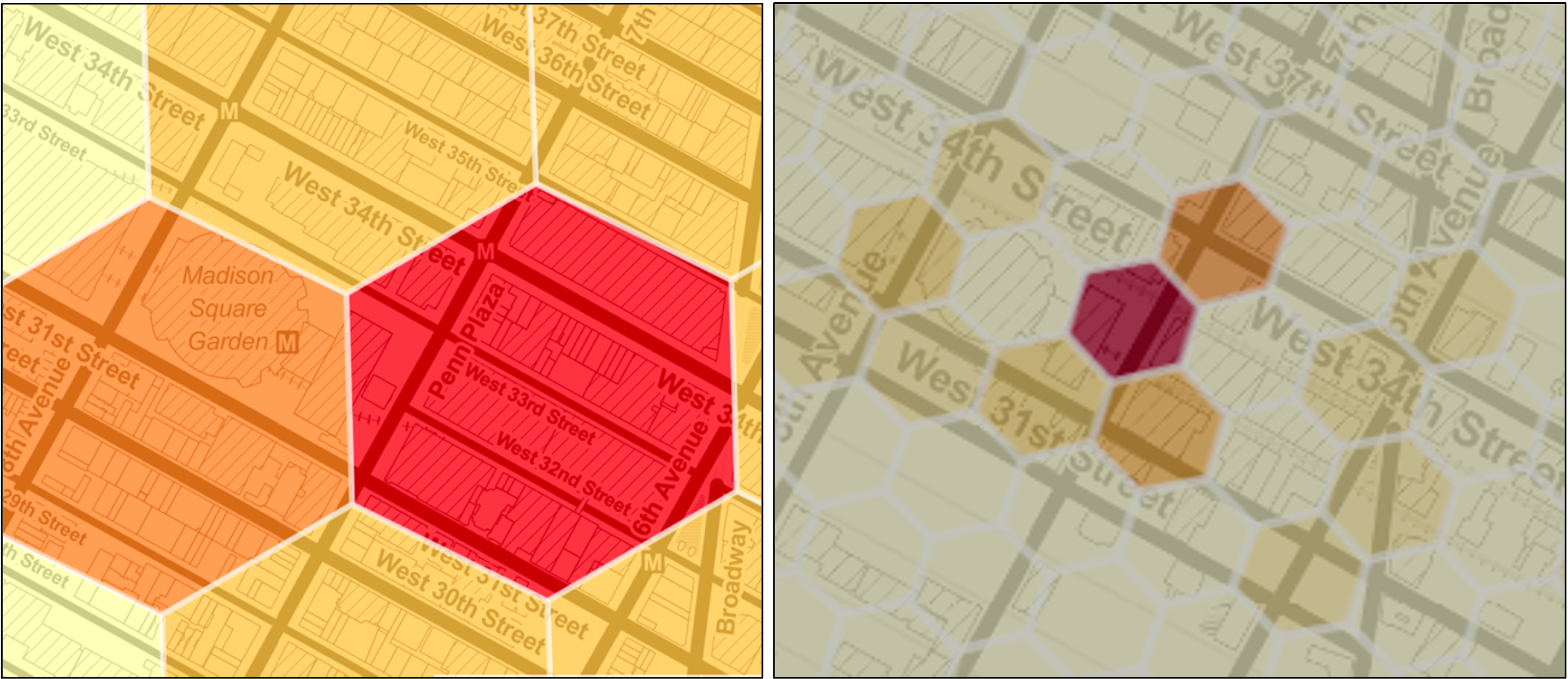}
    \caption{Resolution 9 (left), Resolution 10 (right)}
    \label{fig:resolution}
\end{figure}

The two proposed resolution sizes are resolution 9 (see Figure \ref{fig:resolution} - left), with an average side length or maximum walking distance to the pickup destination of 200 meters, and resolution 10 (see Figure \ref{fig:resolution} - right), with an average side length or maximum walking distance of 75 meters \parencite{h3source}.\par
\begin{figure}
    \centering
    \includegraphics[width = 0.9\textwidth]{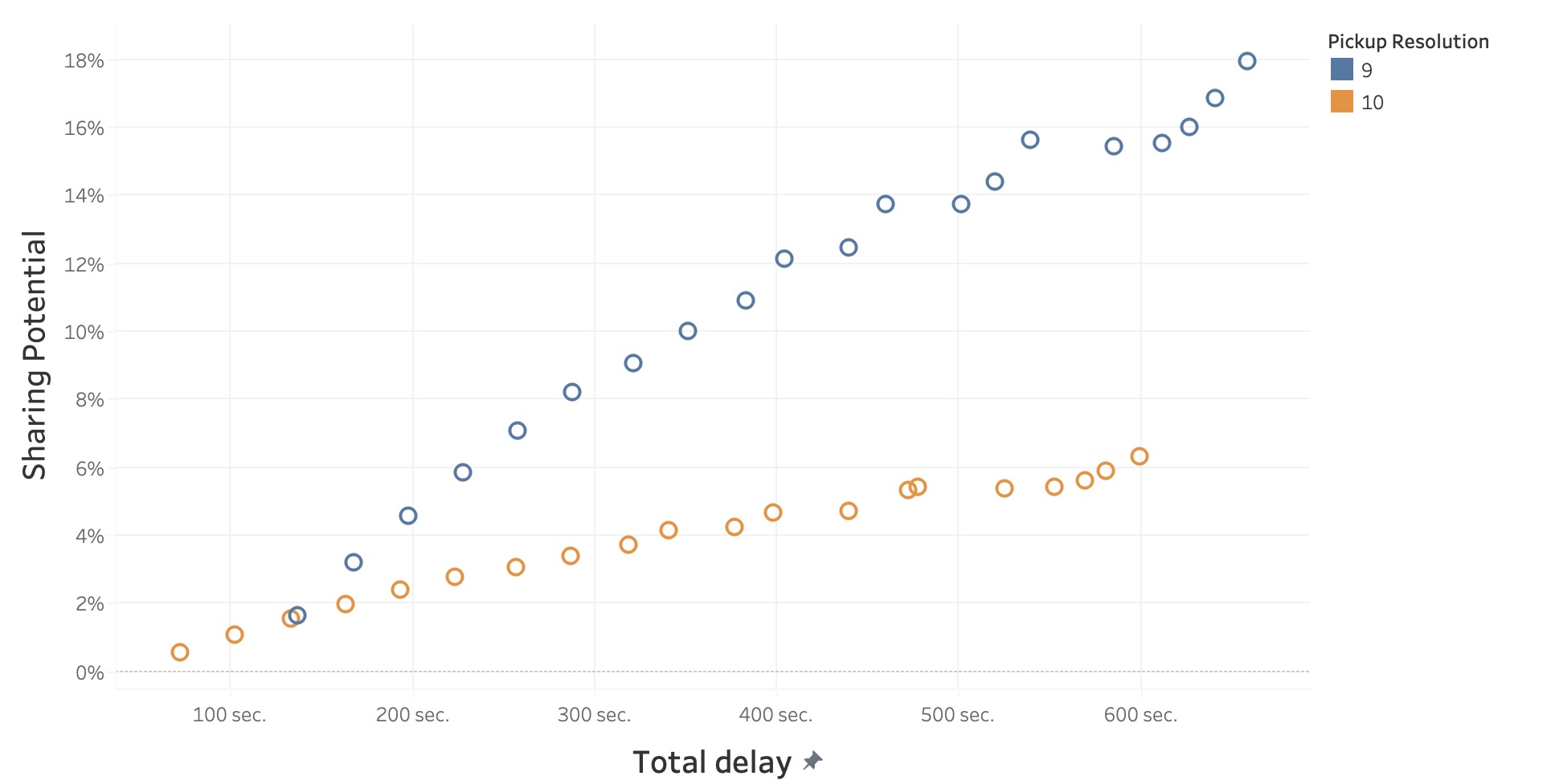}
    \caption{Sharing Potential vs. Total Delay}
    \label{fig:sharing_potential_delay}
\end{figure}

Comparing the results from figure \ref{fig:sharing_potential_delay} with the given time intervals indicates that the resolution size 9 performs better in the ridesharing scenario by still restraining the total delay. Especially with higher time intervals, there is observable that the lower resolution is more beneficial in terms of delay and sharing potential. For example, we compare the average delays at 135 seconds and reckon that the ridesharing potential at resolution 9 is 10\% but only 4\% with a resolution of 10 (see Table \ref{tab:resolution_size_delay}). This supposition becomes even more evident when looking at Figure \ref{fig:sharing_potential_delay}. The bigger hexagon size accounts for more sharing potential per every second of total delay. The only periods where resolution 10 outperforms resolution 9 are below 3 minutes of total delay.

\begin{table}[]
\renewcommand{\arraystretch}{1.2}
\begin{center}
\begin{tabular}{l|llll}
\cline{2-5}
                   & \multicolumn{4}{c|}{Drop-off Resolution}                                                               \\  
                   & \multicolumn{4}{c|}{9}                                                                                                         \\  \cline{2-5}
\multirow{-4}{*}{} & \multicolumn{2}{c|}{Pickup Resolution} &  \multicolumn{2}{c|}{Pickup Resolution}                                                           \\  
\multirow{-4}{*}{} & \multicolumn{2}{c|}{9}    &   \multicolumn{2}{c|}{10}                 \\ \cline{2-5}
T & Sharing Potential & \multicolumn{1}{l|}{Total Delay}
& Sharing Potential & \multicolumn{1}{l|}{Total Delay} \\
1                  & 1.65 \%            & \multicolumn{1}{l|}{137}                                     & 0.57 \%            & \multicolumn{1}{l|}{72}                  \\
2                  & 3.18 \%            & \multicolumn{1}{l|}{168}      & 1.08 \%            & \multicolumn{1}{l|}{102}                 \\
3                  & 4.57 \%            & \multicolumn{1}{l|}{198}                                         & 1.56 \%            & \multicolumn{1}{l|}{133}                 \\
4                  & 5.86 \%            & \multicolumn{1}{l|}{228}                 & 1.98 \%            & \multicolumn{1}{l|}{163}                 \\
5                  & 7.08 \%            & \multicolumn{1}{l|}{258}                                         & 2.39 \%            & \multicolumn{1}{l|}{193}                 \\
6                  & 8.21 \%            & \multicolumn{1}{l|}{288}                 & 2.77 \%            & \multicolumn{1}{l|}{223}                 \\
7                  & 9.07 \%            & \multicolumn{1}{l|}{321}                                         & 3.04 \%            & \multicolumn{1}{l|}{257}                 \\
8                  & 9.98 \%            & \multicolumn{1}{l|}{351}                 & 3.37 \%            & \multicolumn{1}{l|}{287}                 \\
9                  & 10.92 \%           & \multicolumn{1}{l|}{383}                                         & 3.70 \%            & \multicolumn{1}{l|}{319}                 \\
10                 & 12.63 \%           & \multicolumn{1}{l|}{404}                 & 4.12 \%            & \multicolumn{1}{l|}{341}                 \\
11                 & 13.07 \%           & \multicolumn{1}{l|}{440}                                         & 4.24 \%            & \multicolumn{1}{l|}{377}                 \\
12                 & 14.47 \%           & \multicolumn{1}{l|}{460}                 & 4.68 \%            & \multicolumn{1}{l|}{398}                 \\
13                 & 14.66 \%           & \multicolumn{1}{l|}{501}                                         & 4.72 \%            & \multicolumn{1}{l|}{440}                 \\
14                 & 15.47 \%           & \multicolumn{1}{l|}{520}                 & 5.31 \%            & \multicolumn{1}{l|}{472}                 \\
15                 & 16.88 \%           & \multicolumn{1}{l|}{539}                                         & 5.41 \%             & \multicolumn{1}{l|}{478}                 \\
16                 & 16.37 \%           & \multicolumn{1}{l|}{585}                 & 5.36 \%            & \multicolumn{1}{l|}{525}                 \\
17                 & 15.55 \%           & \multicolumn{1}{l|}{612}                                         & 5.42 \%            & \multicolumn{1}{l|}{553}                 \\
18                 & 16.02 \%           & \multicolumn{1}{l|}{627}                 & 5.59 \%            & \multicolumn{1}{l|}{569}                 \\
19                 & 16.83 \%           & \multicolumn{1}{l|}{641}                                         & 5.90 \%            & \multicolumn{1}{l|}{581}                 \\
20                 & 17.92 \%           & \multicolumn{1}{l|}{659}                 & 6.31 \%            & \multicolumn{1}{l|}{599}                 \\ \cline{2-5}
\end{tabular}
\caption{Resolution Size vs. Delay}
\label{tab:resolution_size_delay}
\end{center}
\end{table}

\subsubsection{Trip Intervals}
\begin{figure}
    \centering
    \includegraphics[width = 0.9\textwidth]{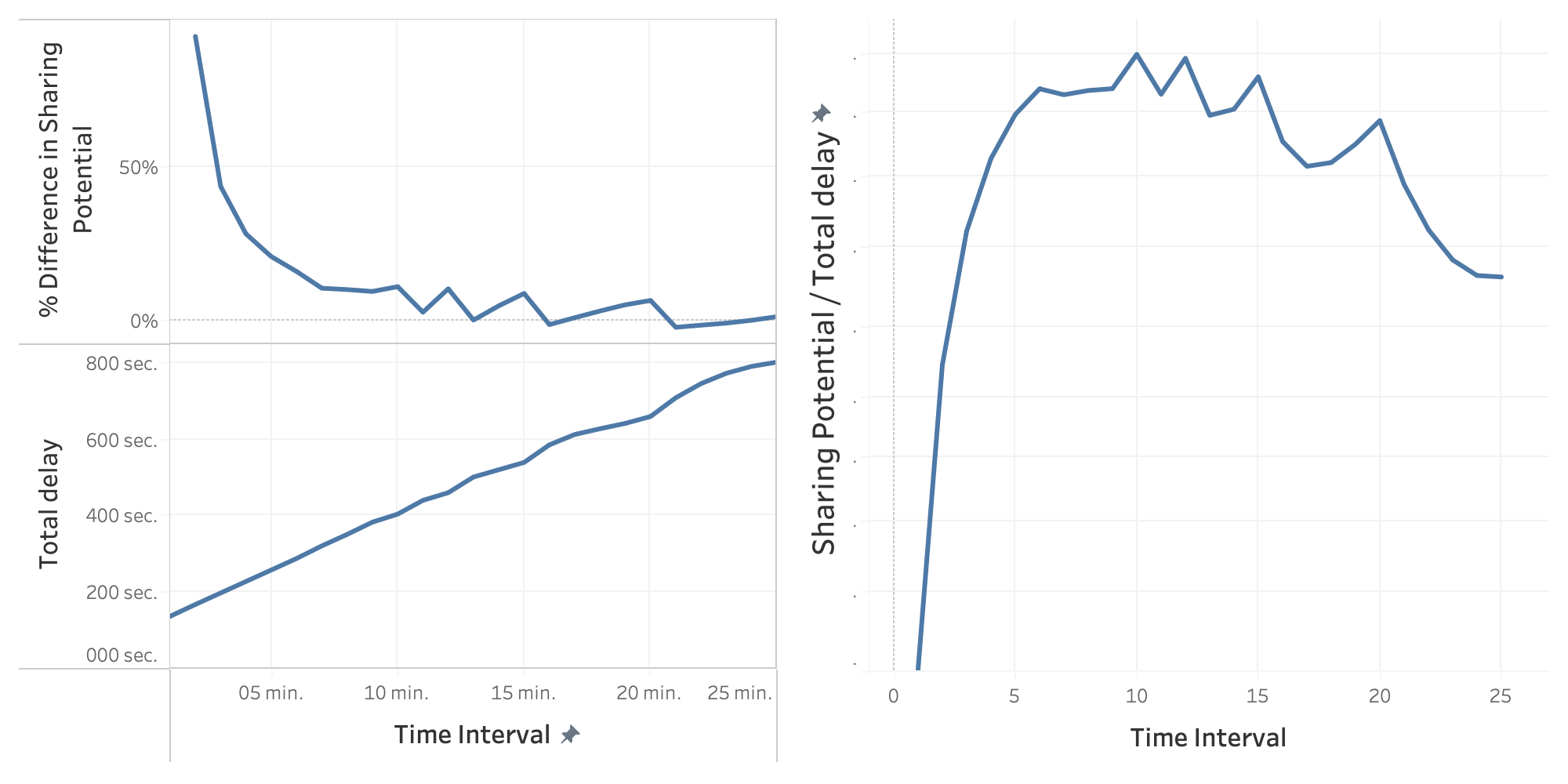}
    \caption{Interval Size vs. Sharing Potential \& Delay}
    \label{fig:inteval_potential_delay}
\end{figure}

In order to evaluate a reasonable time window for subsetting the departure time of all taxis at the hexagon center with resolution 9, Figure \ref{fig:inteval_potential_delay} assesses the growth of the ridesharing potential dependent on the interval size. A first glance reveals that with higher time intervals, the ridesharing potential flattens and moves towards zero growth. Throughout the 25-time intervals, the total delay of shared rides stays almost linear. \par
The percent difference in ridesharing potential follows a steady decrease until interval 10 but then varies in growth from year to year. Following this investigation, we can assume that setting the time interval to 10 and having the pickup resolution at 9 is reasonable. This setting will result in the best sharing potential compared to the total delay (see Figure \ref{fig:inteval_potential_delay} - right).

\subsection{Two-Scenario-Analysis}\label{two-scenario 3.3}
As observable in table \ref{tab:resolution_size_delay}, the trip sharing potential with a resolution of 9 at the pickup and drop-off destination at a given 10-minute time window will result in 12,1\% sharing potential following the presented approach. It is now essential to determine not just the overall average but also analyze how the sharing potential behaves on an ordinary day and extraordinary day \parencite{barann2017open}, as well as figure out how far our findings could be generalized to times with lower and higher taxi demands \parencite{santi2014quantifying}.\par
\begin{figure}
    \centering
    \includegraphics[width = 0.9\textwidth]{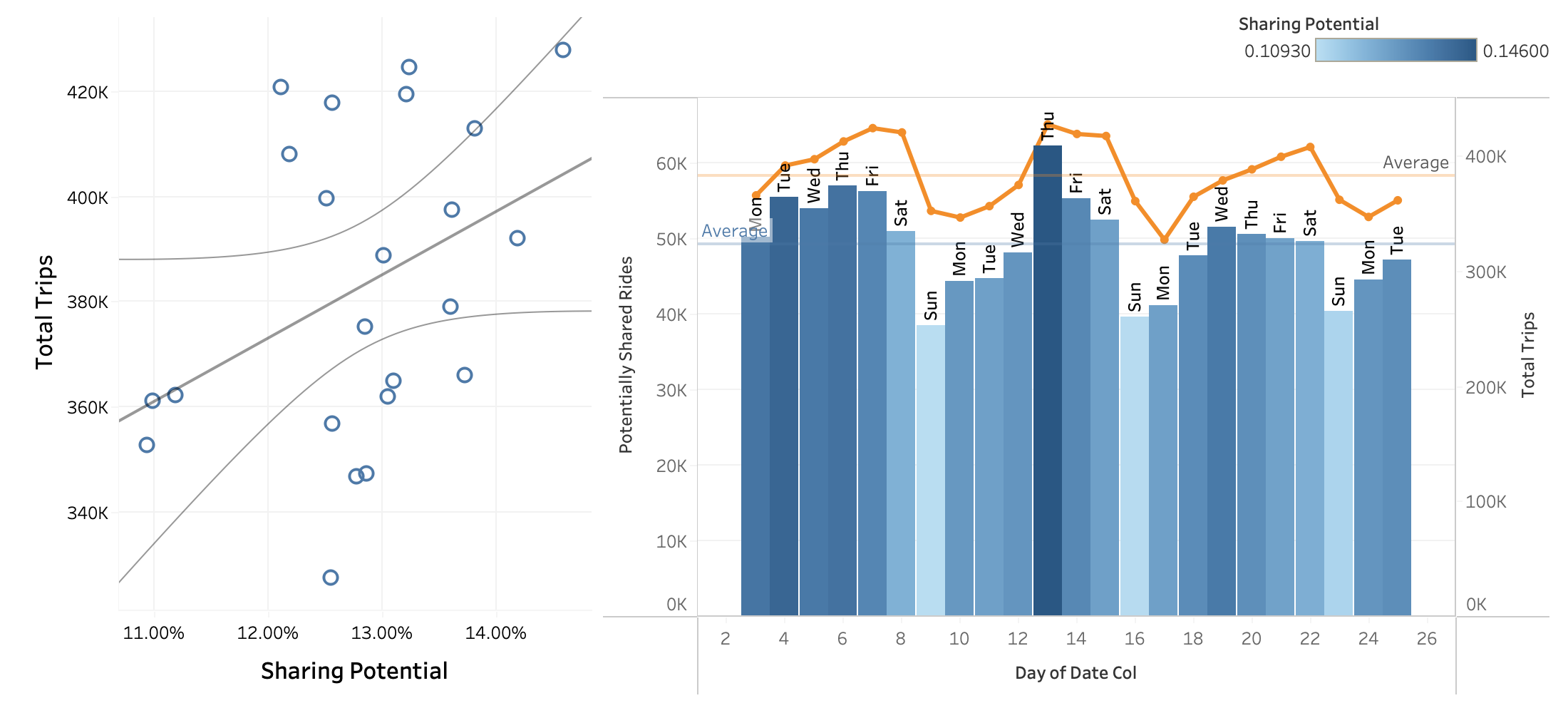}
    \caption{Taxi Demand and Ridesharing Potentialy}
    \label{fig:reg_demand_potential}
\end{figure}

As can be observed in Figure \ref{fig:reg_demand_potential}, there is a clear connection between the number of trips that happened within a day and the potential to participate in ridesharing. The sharp decline in potentially shared rides on the 9th, 16th, and 23rd of March can be traced back to the below-average total trips on these respective days. Another exciting aspect observable in Figure \ref{fig:reg_demand_potential} is that even though the ridesharing potential is lowest on Sundays, it is only the second lowest weekday regarding taxi demand. The three weekdays with the lowest taxi demand in March were all Sundays. Comparing the color spectrum and the size of the bars shows a clear positive correlation between taxi demand and ridesharing potential. The same picture can be seen in other studies (\cite{alonso2017demand},\cite{barann2017open},\cite{santi2014quantifying}).

By comparing all daily taxi demands and ridesharing potentials in March, we
choose an ordinary day based on the average daily demand of 383 thousand trips and the ridesharing potential of 12.9\%. Wednesday, March 20, is likely to represent an ordinary day with an average ridesharing potential of 13\% and a total taxi demand of 388 thousand, whereas Sunday, March 9, represents a minimum sharing potential of 10.9 \% in the observed month and fourth lowest day in terms of taxi demand representing the extraordinary day \parencite{barann2017open}.

\subsubsection{Trip Savings and Environmental Benefits}
As mentioned earlier, 
the trip distance of a shared taxi trip equals the maximum individual trip distance of the pooled passengers in the ridesharing scenario. The analysis reveals that, on average, this approach will save 48.6 \% of the cumulative trip distance in a shared scenario. Examining multi-ridesharing approaches with more than five pooled ride requests, the average cumulative trip distance saved is even 82 \% on both ordinary and extraordinary days. We can estimate the daily environmental emission savings by following these findings and the total distance saved in a day (see table \ref{tab:saved_distance_emissions}). The emission calculation is based on the average emissions of vehicles in the US in 2016, which is approximately 250 grams per kilometer \parencite{EPAGreenhouse}.

\begin{table}[]
\renewcommand{\arraystretch}{1.2}
\begin{center}
\begin{tabular}{lll}
Metrics                    & Ordinary    & Extraordinary \\ \hline
Individual Trips           & 338,099     & 314,066       \\
Shared Trips               & 50,582      & 38,538        \\
Total Reduction of Trips   & 27,865      & 21,205        \\
\%. in vehicle reduction    & 8.24 \%      & 6.75 \%        \\
Total Distance saved       & 80,517 km   & 64,743 km     \\
\% Avg. trip distance saved & 48.56 \%     & 48.63 \%       \\
Saved CO2*                 & 20,129 tons & 16,186 tons  
\end{tabular}
\caption{Saved Distance and Emissions}
\label{tab:saved_distance_emissions}
\end{center}
\end{table}
Within this approach, we assume that there is no capacity limitation. Therefore, some trips must be made with greater capacity vehicles with higher emissions per kilometer. We neglect this fact in our saved CO2 calculation. Nevertheless, knowing what vehicle capacities must be available for the shared ride is crucial. Table \ref{tab:vehicle_types} illustrates how many vehicles of what type must be used to serve the ridesharing demand implementing this approach. For regular cars, a passenger limit of 5 was assumed. We assumed a capacity limit of 10 passengers for vans, and minibusses should transport up to 30 passengers.
\begin{table}[]
\renewcommand{\arraystretch}{1.2}
\begin{center}
\begin{tabular}{lll}
Metrics              & Ordinary & Extraordinary \\ \hline
Car Total Trips      & 17,625   & 13,073        \\
\% Passengers carried & 37.8 \%   & 53.7 \%        \\
Van Total Trips      & 4,631    & 3.746         \\
\% Passengers carried & 55.9 \%   & 37.2 \%        \\
Minibus              & 461      & 514           \\
\% Passengers carried & 6.3 \%    & 9.1 \%        
\end{tabular}
\caption{Vehicle Types}
\label{tab:vehicle_types}
\end{center}
\end{table}

\subsubsection{Delay}
As 
shown in Table \ref{tab:delay}, the level of taxi demand has a negligible effect on total delays. This favors the presented approach since, unlike the research of \parencite{barann2017open}, demand has little effect on delay. The delay is relatively independent of the demands as we suppose a static departure time at the pickup location.
\begin{table}[]
\renewcommand{\arraystretch}{1.2}
\begin{center}
\begin{tabular}{lll}
Metrics                       & Ordinary & Extraordinary \\ \hline
Average Total Delay           & 402 sec. & 404 sec.      \\
Average Walking Duration      & 96 sec.  & 94 sec.       \\
Average walking distance      & 124 m.   & 122 m.        \\
Average Delay until Departure & 390 sec. & 392 sec.      \\
Average Delay at Drop-off     & 10 sec.  & 12 sec.      
\end{tabular}
\caption{Delay}
\label{tab:delay}
\end{center}
\end{table}
\subsubsection{Possible Cost Savings}
The price of a taxi trip is dependent on the length and duration of the trip. This section focuses solely on the potential cost savings passengers could have and neglects that the taxi driver also suffers a delay due to the waiting time at the pickup destination and the additional delay of 20 seconds for every drop-off in the multi-ridesharing situation. Similar to calculating the total Trip distance, we use the maximum fare amount in a ridesharing situation as a foundation and assume that the remaining fares of the ridesharing passengers could be seen as the maximum potential cost savings.
\begin{table}[]
\renewcommand{\arraystretch}{1.2}
\begin{center}
\begin{tabular}{lll}
Metrics                       & Ordinary & Extraordinary \\ \hline
Average cost-saving potential & 49 \%      & 49 \%          
\end{tabular}
\caption{Cost-Saving Potential}
\label{tab:cost_saving_potential}
\end{center}
\end{table}

\subsection{Cost Savings within Airport Trips}\label{Cost Savings Airport 3.4}
As mentioned in the previous 
analysis,
car occupancy is significantly lower with trips to and from the airport. However, the financial incentives for such trips should be higher as the trip fares are relatively costly \parencite{storch2021incentive}. 
Contrary to our 
expectations, the ridesharing potential at the airports JFK and La Guardia is 5 \% below the monthly average of the total analyzed in New York City, showing an 8 \% sharing potential. However, sharing a taxi trip with pickup and drop-off destination at the airports comes with a higher cost benefit, as these trips are, on average, the most extended (see Figure \ref{fig:trip_fare_distance}). Taxis charge an extra fee as they operate at a different rate doing pickups at JFK or La Guardia Airport \parencite{tlcfactbook2014}. In relative figures, the average cost savings in the shared ride scenario at the airport account for 50 \%, similar to the ordinary case demonstrated above. However, on average, a trip from or to the airport in a shared or unshared vehicle costs \$ 42, but a regular trip within Manhattan costs \$ 12.50. Hence, the incentive to split fare costs using ridesharing services is higher for trips from and to the airport than ordinary taxi trips in NYC. 

\subsection{Peak-Hour Scenario}\label{Peak-Hour 3.5}
Next, we study 
how far higher demands could lead to a better performance of the one-to-one sharing approach, and if it is worth implementing a shared service solely available within these high-demand hours. In the peak hour scenario, the observations in low taxi demand hours are (see Figure \ref{fig:peak_hours}) filtered out to assess the sharing potential only considering peak hours of the day. This examination shows that the total ridesharing potential is 13.1\%, given a 10-minute interval. Compared to the base case, this is a negligible increase. Following this, we 
assume that the ridesharing potential is increasing with growing demand but is less drastic than the static one-to-one approach by \parencite{barann2017open}. We can also assess that the average waiting time in such high-demand situations is 390 seconds, 5\% lower than the base case that includes the off-peak hours.

\subsection{Locations with the Highest Sharing Potential}\label{Location Potential 3.6}
\begin{figure}
    \centering
    \includegraphics[width = 0.9\textwidth]{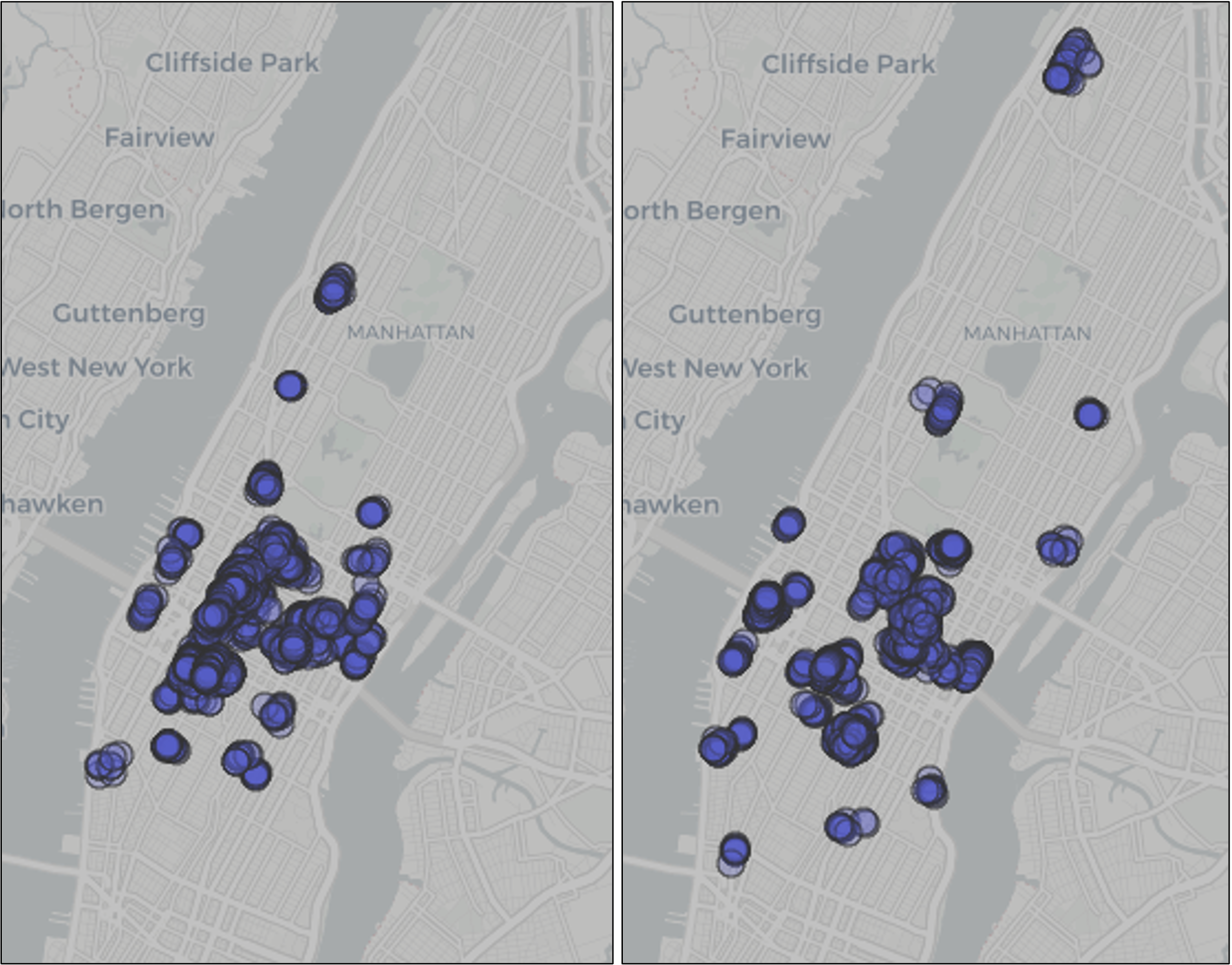}
    \caption{Pickup (left) and Drop-off (right) locations with the highest demand}
    \label{fig:high_demand_loc}
\end{figure}

Exploring all drop-off and pickup locations from March 2014 provides us with a clear picture of destinations with the highest ridesharing potential. The majority of taxi trips are in central Manhattan. However, the best ridesharing opportunities are in southern Manhattan (see Figure \ref{fig:high_demand_loc}). Based on these findings, a Taxi Ride Sharing service in New York City should primarily serve Manhattan's central and southern areas to reach the highest ridesharing potential.\par
A closer look at the pickup hotspots (see Figure \ref{fig:high_demand_loc}) reveals three significant clusters showing the highest ridesharing opportunity: Madison Square Garden, The Grand Central Station, and the intersection at 42nd Street and 8th Avenue. The most prominent clusters for drop-off locations are Madison Square Park and Javits Center for this specific month.\par
In a circumstance where the ridesharing approach is limited to the ten highest-demand pickup areas (visualized in Figure \ref{fig:pickup_demands}), we achieve a matching rate of about 27\% (see table \ref{tab:hot_spot_matching}). Restricting the model to serve these ten hotspots excludes 90\% of New York City's taxi demand. Consequently, this simple approach has the potential to transform 2.5\% of New York taxi rides into a shared service on an ordinary day and around 2\% on an extraordinary day by solely focusing on 10 pickup areas.

\begin{table}[]
\renewcommand{\arraystretch}{1.2}
\begin{center}
\begin{tabular}{lll}
Metrics                       & Ordinary & Extraordinary \\ \hline
Ride Sharing Potential        & 26.7 \%   & 22.9 \%        \\
Representation of total rides & 10.5 \%   & 9.2 \%    
\end{tabular}
\caption{Hotspot Matching Rate}
\label{tab:hot_spot_matching}
\end{center}
\end{table}

\section{Discussion}\label{Discussion}

We have proposed an easy and much-needed sharing strategy 
that pairs trips with similar starting and ending points. 
For a dataset of more than 165 million taxi rides,
we have shown that the cumulative journey time can be reduced by 48 percent,
fares can be reduced by 49 percent,
while emissions can be decreased by 20.129 tons of CO2 -- on an ordinary day.
Compared to many-to-many sharing dynamic routing schemes,
our scheme is substantially easier to implement and operate.

We have demonstrated that a ridesharing provider 
can adjust  to the complex distance and matching time constraints simultaneously and how to do so. 
While the more patient a passenger is, the more significant the matching opportunities that occur. Details on passenger behaviors may vary and are often scarce,
making it difficult to develop a robust framework.
We have revealed that reducing the matching time interval has a significant detrimental effect on the matching rate. Therefore, one may argue that a 10-minute time frame may be optimal for a taxi ridesharing service in New York City, whereas at other places someone else would never want to get delayed by 400 seconds to save up to 49\% of the trip costs.
Nevertheless, the time limitation mitigates the effect of the distance constraint, and its arrangement also determines a fair distribution of disadvantages of the passengers in a shared trip. 
While the distance limitation increases, the resolution size decreases, and 
distant passengers must walk to join a transport \parencite{barann2017open}.\par
Our analysis robustly demonstrates that a varying trip density does not
render our one-to-one approach inefficient.
A lower trip density results in a lower matching rate, but considering the resulting 13 \% matching rate on an ordinary day, a 10.9 \% matching rate on a low trip demand day, and a 13.1 \% matching rate only considering high demand hours demonstrates the robustness of our conclusions.
In addition, our approach is more durable than a previously suggested one-to-one model \parencite{barann2017open}. Their approach 
showed 53 \% ridesharing potential in high demand but only 21\% in low demand times. Nonetheless, as the number of ride requests increases, the trip density's impact on the matching rate diminishes.
In fact, 
we achieve a
matching rate of 27\% 
on an ordinary day by concentrating on the local pickup hotspots. 
Yet, 
information on spatial and temporal characteristics of such high-demand trips and matching situations may 
be exploited 
for
implementing a public transportation service with multi-ridesharing potential.

\parencite{santi2014quantifying} demonstrated that a flexible, dynamic ridesharing system may achieve close to maximum matching by using 25\% of daily cab journeys in New York City, or around 100,000 trips per day. Therefore, they concluded that ridesharing is also effective in low-density cities. \par
In contrast, \parencite{stiglic2015benefits} concluded that the ridesharing potential in traditional ridesharing depends heavily on the density of spatial and temporal distribution of the stated trips \parencite{barann2017open}.\par
According to related studies, there is a low likelihood of passengers wanting to adopt ridesharing, even if they are financially incentivized. \parencite{stiglic2015benefits} demonstrated that ridesharing initiatives may fail. 
Understanding this 
is ultimately important and particularly applicable to our suggested one-to-one ridesharing strategy, as passengers must wait for possible mates and have the time to decide until departure if they want to share the ride, especially considering that the passengers can see with whom they would share the ride. 
Our 
analysis suggests 
that the proportion of matched rides in New York City 
may reach up to 13\%, providing 
evidence that the proposed strategy can be successfully applied. 
As we have mapped out certain high-demand areas and hotspots for pickups and drop-offs, one could implement a ridesharing service in these areas that performs at a higher rate.\par
Taxi firms should consolidate shared journeys using hotspots, such as transit hubs or notable landmarks, as starting and ending sites \parencite{furuhata2013ridesharing}. 
Models suggest 
that picking meeting locations with care can considerably enhance the number of matched rides in traditional ridesharing scenarios \parencite{stiglic2015benefits}. 
Our findings indicate that multimodal transportation hubs such as 
Madison Square Garden could be 
ideal 
for a one-to-one sharing service. Also, considering airports as shared-ride pickup hotspots due to the high financial incentives, they may efficiently 
help increase 
adoption even though the matching success is lower than in Manhattan.
\printbibliography
\end{document}